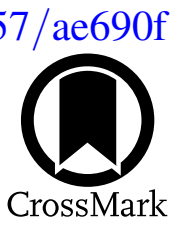

# Evidence for Dynamical Filtering: High Binary Fraction, Hard-binary Excess, and Unresolved Triples in the Surviving Core of NGC 6791

Huanbin Chi (迟焕斌)[1,2,3,4], Zhi Li (李志)[5,6], Feng Wang (王锋)[3,4], Xuefen Tian (田雪芬)[2], Linfeng Chang (常林凤)[2], Hongbo Liu (刘红波)[1], and Yiqin Liu (刘艺琴)[2]
[1] Institute of Artificial Intelligence Applications, Yunnan Open University, Kunming 650599, People's Republic of China
[2] School of Artificial Intelligence, Yunnan Open University, Kunming 650599, People's Republic of China
[3] Center For Astrophysics, Guangzhou University, Guangzhou, Guangdong 510006, People's Republic of China; fengwang@gzhu.edu.cn
[4] Great Bay Center, National Astronomical Data Center, Guangzhou, Guangdong 510006, People's Republic of China
[5] Yunnan Observatories, Chinese Academy of Sciences, Kunming 650216, People's Republic of China
[6] Key Laboratory for Structure and Evolution of Celestial Objects, Chinese Academy of Sciences, People's Republic of China

## Abstract

We present a deep photometric analysis of the main-sequence (MS) population in the old, metal-rich open cluster (OC) NGC 6791 using Gaia Data Release 3 data. After correcting for differential reddening, we use the Bayesian model comparison to test whether stellar rotation can account for the observed MS broadening and find that a rotation-dominated interpretation is strongly disfavored. We therefore infer that unresolved multiplicity is the primary contributor to the photometric offsets. We derive a high-$q$ companion fraction of 54.3% ± 2.8% for systems with $q \gtrsim 0.5$, significantly higher than typical values reported for most OCs and the field. The inferred offset distribution is not consistent with a flat mass-ratio distribution but instead shows an excess toward high mass ratios ($q \sim$ 0.8–1.0), suggestive of preferential survival of hard binaries in a dynamically evolved environment. We also identify a population of stars lying above the equal-mass binary limit ($\Delta G > 0.75$ mag), which is difficult to explain with ordinary MS binaries alone and is plausibly interpreted as candidate unresolved triple or higher-order multiple systems. A Kolmogorov–Smirnov test, together with Monte Carlo label-shuffling experiments, shows no statistically significant difference between the projected radial distributions of the single-star and binary/multiple populations within the observed field. Taken together, these results are consistent with the picture that NGC 6791 is the dynamically processed inner remnant of a once more massive cluster.

## 1. Introduction

Binary stars are not merely passive tracers of star formation; in the dense environments of star clusters, they act as fundamental engines of dynamical evolution. Through three-body interactions, binary systems exchange energy with passing single stars. According to the foundational Hills–Heggie law (D. C. Heggie 1975), these interactions bifurcate the binary population: "soft" binaries (weakly bound, typically with low mass ratios $q$) tend to be disrupted, while "hard" binaries (strongly bound) become tighter, releasing kinetic energy that heats the cluster core and halts core collapse (P. Hut et al. 1992; S. F. Portegies Zwart et al. 2010). Consequently, the present-day binary fraction ($f_b$) and the mass-ratio distribution ($f(q)$) of an evolved cluster serve as dynamical tracers of the cumulative effects of stellar encounters, evaporation, and mass segregation (A. Sollima et al. 2010; A. P. Milone et al. 2012). While globular clusters have been extensively studied in this context, old open clusters (OCs) offer a unique, intermediate density regime to test these theories, yet few survive beyond a few gigayears due to disruption by the Galactic tidal field.

NGC 6791 stands as a paramount exception. It is one of the oldest (∼8.0–8.3 Gyr) and most metal-rich ([Fe/H] ≈ +0.3 to + 0.4) OCs known in the Milky Way (F. Grundahl et al. 2008; K. Brogaard et al. 2012; S. T. Linden et al. 2017). Located at a distance of $d \approx 4.0$ kpc and a height of $Z \approx 0.8$ kpc above the Galactic plane, it occupies an orbit that has been argued to reduce the impact of disk shocking relative to typical solar-neighborhood OCs (G. Carraro et al. 2006; L. A. Martinez-Medina et al. 2018; S. Villanova et al. 2018; I. L. Colman et al. 2022; V. V. Jadhav et al. 2023; N. M. Ahmed 2025). Recent dynamical studies suggest that the presently observed cluster is well contained within its estimated Jacobi radius ($R_{\rm J}$), with extremely small ratios of half-mass and tidal radii to $R_{\rm J}$ ($r_h/R_{\rm J}$ and $r_t/R_{\rm J}$), indicating a system that is dynamically relaxed and dominated by internal gravity rather than external tides (N. Alvarez-Baena et al. 2024). Here the Jacobi radius, $R_{\rm J}$, denotes the effective tidal boundary of the cluster, where the cluster's self-gravity and the Galactic tidal field are approximately in balance. With a present-day mass exceeding 4000 $M_\odot$ (I. Platais et al. 2011), NGC 6791 is likely the remnant core of a much more massive initial system ($\sim 10^5\, M_\odot$) that has lost its outskirts to evaporation, making it a useful laboratory for studying dynamical filtering of the stellar population (E. Dalessandro et al. 2015).

The color–magnitude diagram (CMD) of NGC 6791 has long puzzled astronomers due to the significant broadening of its main sequence (MS). While MS broadening is a classic signature of unresolved binaries, it is degenerate with other physical effects, primarily differential reddening (due to variable dust extinction across the field), metallicity spread, and rapid stellar rotation (N. Bastian & S. E. de Mink 2009; T. D. Brandt & C. X. Huang 2015). Previous high-resolution imaging with the Hubble Space Telescope (L. R. Bedin et al. 2008) suggested a binary fraction of ∼30%, but limited fields of view and the lack of proper motion membership hampered a comprehensive census. More recently, high-resolution spectroscopy has ruled out significant metallicity spreads ($\sigma_{[\mathrm{Fe/H}]} < 0.02$ dex; S. Villanova et al. 2018), leaving rotation and multiplicity as the primary competing explanations.

The release of Gaia Data Release 3 (DR3; Gaia Collaboration et al. 2023) provides the precise photometry and astrometry needed to break these degeneracies. Although N. M. Ahmed (2025) recently utilized Gaia DR3 to analyze the fundamental parameters and blue stragglers of NGC 6791, their study focused primarily on the global characterization of the cluster. A dedicated, high-resolution study focusing on the detailed mass-ratio distribution of the lower MS is still lacking. Our work distinguishes itself from previous studies by (1) applying a Bayesian framework to test whether stellar rotation can account for the observed MS broadening, (2) providing a revised estimate of the photometric high-$q$ companion fraction, and (3) identifying a population of candidate unresolved higher-order multiple systems in the cluster core. By doing so, we aim to use the observed binary demographics as tracers of the cluster's long-term dynamical evolution; and (4) the role of hierarchical triples—which are theoretically predicted to form efficiently in high-density cores via binary–binary interactions (N. W. C. Leigh & A. M. Geller 2013)—remains poorly quantified in this cluster.

In this paper, we present a deep photometric analysis of the MS population in NGC 6791. Our work distinguishes itself from previous studies by (1) applying a rigorous Bayesian framework to strictly exclude stellar rotation as the cause of MS broadening, (2) providing a revised, contamination-free measurement of the binary fraction $f_b(q)$, and (3) reporting the first quantitative evidence for a significant population of unresolved triples in the cluster core. We show that the binary population is not only exceptionally abundant but also strongly skewed toward high mass ratios, consistent with a scenario of intense dynamical processing and single-star evaporation.

The paper is organized as follows: Section 2 describes the data reduction, membership selection, and differential reddening correction. Section 3 details our Bayesian methodology for distinguishing between rotation and multiplicity. Section 4 presents the derived binary fraction, mass-ratio distribution, and evidence for triples. Finally, we discuss the dynamical implications of our findings in Section 5.

## 2. Data and Membership Determination

### 2.1. Supervised Machine Learning Identification

We utilized high-precision astrometry and photometry from the Gaia DR3 (Gaia Collaboration et al. 2023). To construct a robust and highly complete member catalog for NGC 6791, we developed a two-step hybrid pipeline combining supervised machine learning with rigorous kinematic pruning, replacing traditional unsupervised clustering methods inspired by H. Chi et al. (2023).

First, we implemented an Extreme Gradient Boosting algorithm (XGBoost; T. Chen & C. Guestrin 2016) to identify candidate members. To construct the training set, we cross-matched our Gaia DR3 catalog (within a $1^{\circ}$ radius of the cluster center) with the highly reliable cluster census provided by E. L. Hunt & S. Reffert (2024). Stars with membership probabilities $>0.5$ in the E. L. Hunt & S. Reffert (2024) catalog were adopted as positive labels (cluster members). For negative labels (field stars), we randomly sampled stars located outside 0.8 times the cluster radius to represent the background/foreground contamination. To enhance the model's discriminative power, the negative-to-positive sample ratio was set to 30:1.

The XGBoost classifier was trained on a six-dimensional feature space comprising both astrometry and photometry: proper motions ($\mu_{\alpha*}$, $\mu_{\delta}$), parallax ($\varpi$), $G$-band mean magnitude, $G_{\mathrm{BP}} - G_{\mathrm{RP}}$ color, and total proper motion ($\mu_{\mathrm{total}}$). We utilized the `Optuna` framework (T. Akiba et al. 2019) to automatically optimize hyperparameters (e.g., `n_estimators`, `max_depth`, `learning_rate`) via threefold cross validation, minimizing the log-loss metric. Applying the optimized model to the entire field, we initially selected candidate members with an aggressive machine learning probability threshold of $P_{\mathrm{ML}} > 0.9$.

### 2.2. Kinematic Refinement via Mahalanobis Distance

While the ML model efficiently separates the sequence in the CMD, field stars with similar photometry can still contaminate the sample. To ensure absolute kinematic purity, we performed a strict refinement based on the 3D astrometric space ($\mu_{\alpha*}$, $\mu_{\delta}$, $\varpi$).

We fitted a single-component Gaussian mixture model to the astrometric distribution of the high-probability candidates to determine the intrinsic kinematic center and the covariance matrix ($\Sigma$) of the cluster. For each candidate star, we calculated the Mahalanobis distance ($D_M$), defined as

$$D_M = \sqrt{(\boldsymbol{x} - \boldsymbol{\mu})^T \Sigma^{-1} (\boldsymbol{x} - \boldsymbol{\mu})}, \tag{1}$$

where $\boldsymbol{x}$ is the astrometric vector of the star and $\boldsymbol{\mu}$ is the cluster's kinematic center. Since the astrometric parameters approximately follow a multivariate normal distribution, $D_M^2$ follows a chi-square ($\chi^2$) distribution with 3 degrees of freedom. We applied a rigorous $3\sigma$ threshold based on the cumulative distribution function of the $\chi^2$ distribution, discarding any candidates that fell outside this kinematic boundary.

As demonstrated in Figure 1, our hybrid pipeline yields a final census of 2657 high-confidence members. Compared to the 2062 members identified by E. L. Hunt & S. Reffert (2024), our method successfully expands the cluster's spatial and radial coverage while maintaining an exceptionally tight vector point diagram and a razor-thin MS.

### 2.3. Differential Reddening and Sample Definition

To mitigate the photometric broadening caused by spatially variable extinction, we applied a differential reddening correction following the local-neighbor approach described by A. P. Milone et al. (2012) and adapted by H. Chi & F. Wang (2025; details are provided in Appendix A).

For the core analysis of the binary population, we restricted our sample to the deep MS range of $17.4 < G < 18.8$. Based on the best-fit PARSEC isochrone (see Section 3), this magnitude range primarily samples unevolved MS stars with masses between approximately 0.85 and $1.0\,M_{\odot}$. This strict boundary avoids the evolutionary degeneracy present at the turn-off and subgiant branch (SGB). This selection minimizes evolutionary degeneracies near the turn-off and SGB, allowing the photometric offset $\Delta G$ to be used as an approximate proxy for the binary mass ratio $q$.

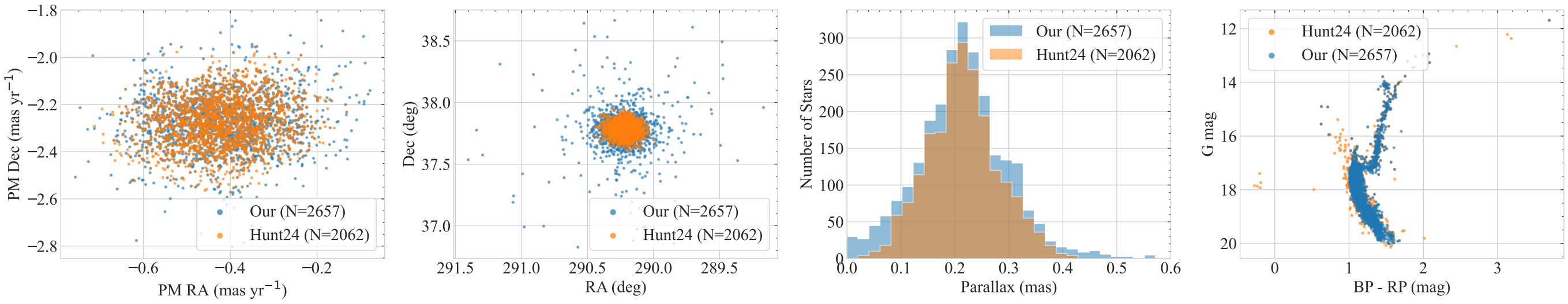

**Figure 1.** Comparison between our kinematically refined member catalog (red, $N$ = 2657) and the catalog from E. L. Hunt & S. Reffert (2024; blue circles, $N$ = 2062). Left: vector point diagram demonstrating the tight proper motion core. Middle: spatial distribution showing our expanded coverage. Right: extinction-corrected CMD highlighting the well-defined cluster sequence obtained after membership selection.

### 2.4. Differential Reddening, Extinction, and Photometric Uncertainties

The line of sight toward NGC 6791 is affected by nonnegligible extinction, and spatially variable reddening can in principle broaden the observed MS. For the absolute placement of the PARSEC isochrone in the Gaia CMD, we adopt the literature reddening toward NGC 6791 and transform the extinction into the Gaia passbands using a standard $R_V = 3.1$ extinction law. Specifically, we use the coefficients $A_G = 0.836 A_V$, $A_{\rm BP} = 1.08 A_V$, and $A_{\rm RP} = 0.63 A_V$. These coefficients are used only to place the fiducial isochrone in the observed CMD; the local spatial variation of the reddening is corrected empirically from the cluster sequence itself.

To assess the possible impact of differential reddening, we applied a local-neighbor correction following the approach of A. P. Milone et al. (2012). In brief, local color residuals relative to the MS ridge line are estimated from nearby high-probability cluster members on the sky, and each star is shifted along the reddening vector by the median local offset. In the magnitude interval adopted for the binary analysis, $17.4 < G < 18.8$, the MS color dispersion changes from 0.0695 mag before the correction to 0.0662 mag after the correction, corresponding to a reduction of 0.0033 mag, or 4.7%. The modest size of this reduction indicates that spatially coherent differential reddening is not the dominant source of the residual MS width in this corrected sample.

We also quantified the Gaia DR3 photometric uncertainties in the same magnitude range. The median uncertainty in the $G$ band is $\sigma_G = 0.0010$ mag, while the median color uncertainty is $\sigma_{\rm BP-RP} = 0.0140$ mag. The $G$-band uncertainty is therefore negligible compared with both the adopted binary threshold, $\Delta G = 0.15$ mag, and the equal-mass binary limit, $\Delta G \simeq 0.75$ mag. The color uncertainty can introduce a finite uncertainty in the inferred vertical offset because $\Delta G$ is measured relative to the MS ridge line at a given color. For a local MS slope $S = dG/d({\rm BP} - {\rm RP})$, the corresponding contribution is approximately $S\sigma_{\rm BP-RP}$, which remains far smaller than the $\Delta G > 0.75$ mag threshold for candidate higher-order multiples.

We therefore treat stars close to the $\Delta G = 0.15$ mag boundary with appropriate caution, since residual color scatter may affect the classification of objects near this threshold. However, random Gaia photometric errors and residual differential reddening are far too small to move ordinary single stars or equal-mass binaries into the $\Delta G > 0.75$ mag region in significant numbers. The high-offset tail is thus unlikely to be an artifact of dust or photometric scatter alone.

## 3. Methods

### 3.1. Isochrone Fitting and Stellar Models

To model the cluster sequence, we employed PARSEC v2.0 isochrones (A. Bressan et al. 2012; C. T. Nguyen et al. 2022). The best-fitting isochrone, determined by $\chi^2$ minimization on the MS, corresponds to an age of 8.0 Gyr, metallicity [Fe/H] = +0.30, distance modulus $(m - M)_0 = 13.05$ (distance $\approx$ 4.1 kpc), and extinction $A_G = 0.65$ mag, which is slightly higher than traditional values (e.g., for S. Bijavara Seshashayana et al. 2025, an $A_V$ of 0.7 corresponds to an $A_G$ of 0.585) but yields the best photometric fit for our specific deep MS region and assuming an extinction law (J. A. Cardelli et al. 1989) with $R_V = 3.1$, giving $E({\rm BP} - {\rm RP}) = 0.416 A_V$ and $A_G = 0.836 A_V$. We note a slight mismatch between the isochrone and the observed SGB, a known challenge in fitting old, metal-rich populations. Our fitting procedure was deliberately weighted to prioritize the populous MS, as our subsequent binary analysis is focused exclusively on MS stars where the fit is well (see Figure 2).

### 3.2. Bayesian Exclusion of Rotation as a Broadening Mechanism

While magnetic braking is expected to have significantly decelerated the rotation of stars in an 8 Gyr old cluster like NGC 6791, we explicitly include a rotation model to quantify any residual degeneracy and ensure that the observed MS broadening is not biased by anomalous stellar rotation.

Before quantifying binaries, we first assess whether rapidly rotating stars could make a significant contribution to the observed MS broadening. We constructed two synthetic cluster models using the PARSEC isochrones: one without rotation ($M_{\rm no_rot}$) and one incorporating rapid rotation ($M_{\rm rot}$) by allowing a fraction $f_{\rm fast}$ of stars to be placed on a rotationally widened isochrone (with $\omega/\omega_{\rm crit} > 0.5$, assuming a uniform distribution of inclinations and a simple color-dependent magnitude offset calibrated on known fast rotators in the Pleiades; K. R. Covey et al. 2016). Both models include only single stars (no binaries) and are convolved with the nominal Gaia DR3 photometric uncertainties.

We performed a Bayesian model comparison using nested sampling as implemented in `dynesty` (J. S. Speagle 2020). The likelihood for a given star is computed by summing over the isochrone points weighted by a Salpeter initial mass function (IMF; $dN/dm \propto m^{-2.35}$) and a Gaussian noise model with variance $\sigma^2 = (0.05\ {\rm mag})^2$, which accounts for both photometric errors and intrinsic model imperfections. The log-likelihood for the entire dataset is the sum over individual stars. We used uniform priors: $\log_{10}(\text{age/yr}) \in [9.0, 10.1]$, $[\text{Fe/H}] \in [0.006, 0.4]$, distance $d \in [1000, 5000]$ pc, $A_V \in [0, 1]$ mag, and for the rotation model, $f_{\text{fast}} \in [0, 1]$. Nested sampling was run with 300 live points, a stopping criterion of $\Delta \ln Z = 0.5$, and the "multi" bounding option with "rwalk" sampling.

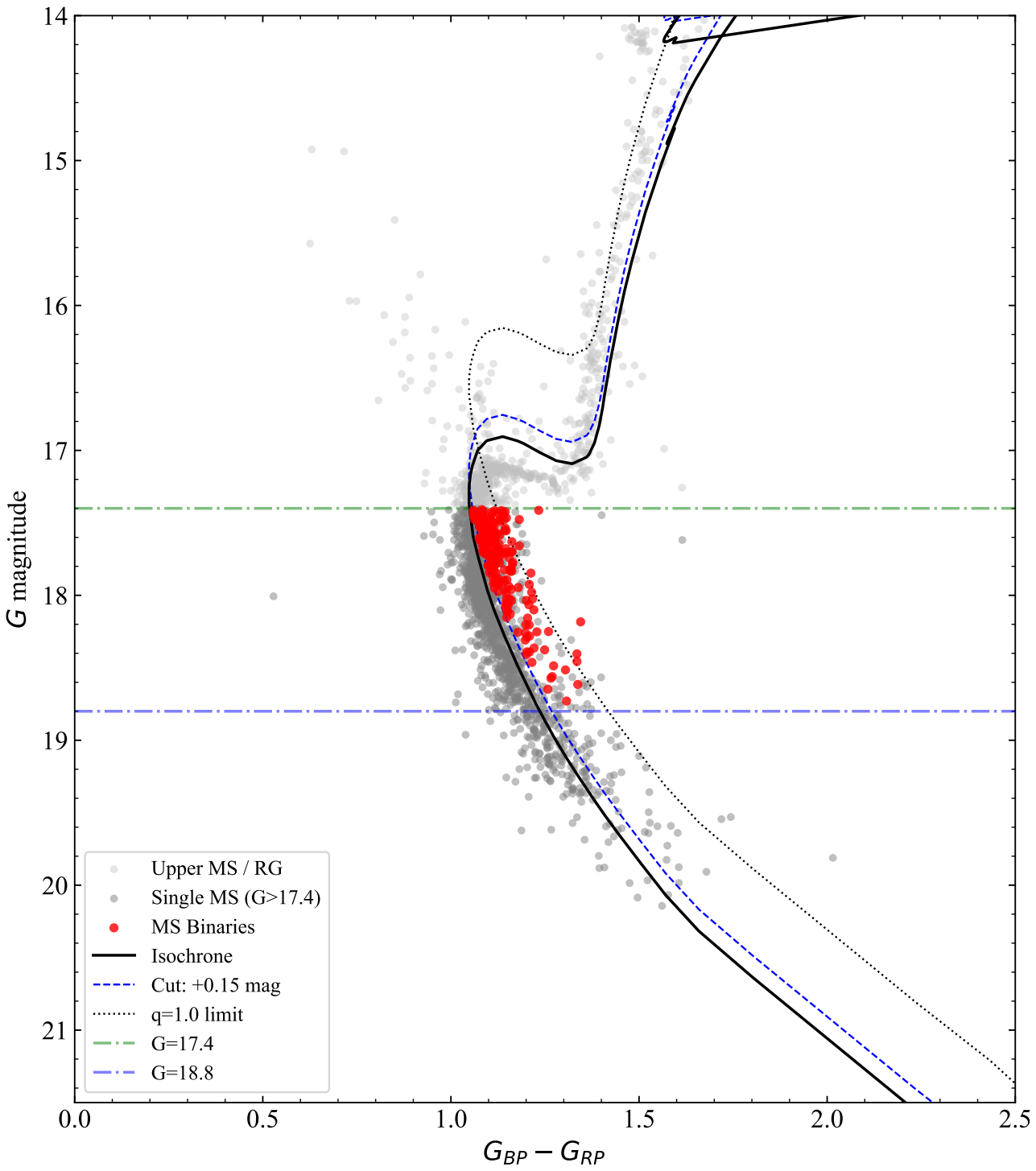

**Figure 2.** CMD of NGC 6791 after differential reddening correction. The solid line represents the fiducial isochrone. The shaded regions illustrate our selection criteria for single stars (gray), binaries (red), and candidate triples (red).

The resulting log-evidence for the rotation model was $\ln Z_{\text{rot}} = -1123.4 \pm 0.3$ and for the no-rotation model $\ln Z_{\text{no_rot}} = -1126.5 \pm 0.3$, yielding a Bayes factor $\ln K = \ln Z_{\text{rot}} - \ln Z_{\text{no_rot}} = -3.1 \pm 0.4$. According to the Jeffreys scale, this constitutes strong evidence against the rotation-dominated model. Furthermore, the posterior distribution for $f_{\text{fast}}$ is consistent with zero ($f_{\text{fast}} = 0.12 \pm 0.12$). We therefore conclude that rapid rotation cannot explain the observed MS width. Given the cluster's confirmed chemical homogeneity, we infer that the photometric offsets must be overwhelmingly dominated by unresolved multiplicity.

### 3.3. Mapping ΔG to Mass Ratio q

We define the vertical photometric offset, $\Delta G$, as the magnitude difference between a star and the fiducial single-star ridge line at a constant color. For an unresolved binary with mass ratio $q = M_2/M_1$ (assuming $M_1 \geqslant M_2$), the flux ratio is approximately $F_2/F_1 = q^{\alpha}$, where $\alpha$ is the exponent of the mass–luminosity relation ($L \propto M^{\alpha}$). For solar-type MS stars, $\alpha \approx 3.5$–4.0 (e.g., Z. Eker et al. 2015). The total magnitude difference relative to a single star of mass $M_1$ is then

$$\Delta G \approx -2.5 \log_{10}(1 + q^{\alpha}). \tag{2}$$

This relation is monotonic in $q$, allowing us to use $\Delta G$ as an approximate proxy for mass ratio over the restricted MS interval considered here. For $\alpha = 3.8$, $\Delta G = 0.15$ mag corresponds to $q \approx 0.5$, and $\Delta G = 0.75$ mag corresponds to $q = 1$ (equal-mass binary). Values substantially exceeding 0.75 mag are difficult to reproduce with ordinary MS binaries alone and are therefore strong candidates for triple- or higher-order systems.

We adopt an operational detection threshold of $\Delta G_{\text{min}} = 0.15$ mag. This threshold is much larger than the median $G$-band photometric uncertainty in the adopted magnitude range, but objects close to this boundary may still be affected by residual color scatter and uncertainties in the ridge-line projection. The resulting fraction should therefore be interpreted as a photometric high-$q$ companion fraction under our adopted CMD selection rather than as a fully deconvolved intrinsic binary fraction. This conservative cut minimizes false positives due to noise while remaining sensitive to binaries with $q \gtrsim 0.5$.

Based on these definitions, we classify the population into three zones (see Figure 2):

1. *Single stars*. $\Delta G < 0.15$ mag. This includes true singles and low-$q$ binaries unresolved by photometry.
2. *MS binaries*. $0.15 < \Delta G \leqslant 0.75$ mag. This range captures systems with mass ratios $q \gtrsim 0.5$ up to equal-mass binaries.
3. *Candidate multiples*. $\Delta G > 0.75$ mag. Objects in this "forbidden zone" strongly suggest the presence of unresolved triples or higher-order systems.

### 3.4. Contamination Estimate for Triple Candidates

To assess whether the $\Delta G > 0.75$ mag stars could be chance alignments with unrelated field stars, we estimated the expected number of contaminants. Using the Besançon Galactic model (A. C. Robin et al. 2003), we simulated the stellar density in the direction of NGC 6791. Assuming a uniform Poisson distribution of field stars with a surface density $\Sigma_{\text{field}}$, the probability of a chance alignment within an angular radius $r$ is

$$P \approx \Sigma_{\text{field}} \pi r^2. \tag{3}$$

For typical Galactic field densities ($\Sigma_{\text{field}} \sim 10^{-6}$–$10^{-5}$ arcsec$^{-2}$), the probability of finding an unrelated star within 1″ is therefore $P \sim 10^{-5}$. The probability of finding an unrelated star within a 1″ radius of a cluster member is $\sim 10^{-5}$. This estimate is consistent with previous contamination analyses in stellar cluster studies, where the field-star surface density is several orders of magnitude lower than the cluster core density. Given our restricted MS analysis sample size of $N \simeq 1353$ stars in the range $17.4 < G < 18.8$, this corresponds to an expected number of chance-alignment contaminants of $\lesssim 0.02$, far below the observed number of high-$\Delta G$ candidates.

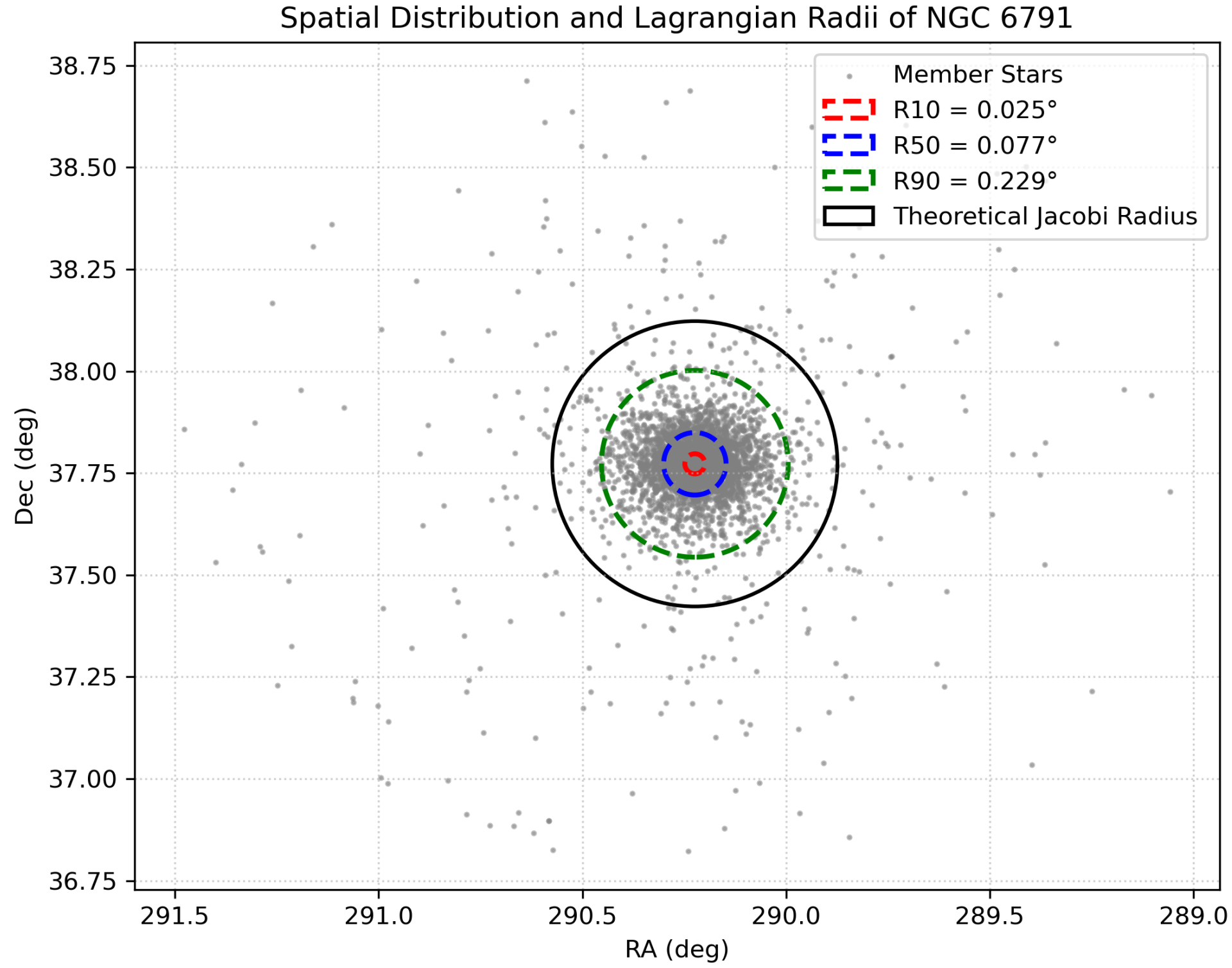

**Figure 3.** Spatial distribution of the identified cluster members in NGC 6791. The concentric dashed lines represent the Lagrangian radii $R_{10}$ (red), $R_{50}$ (blue), and $R_{90}$ (green). The outermost black solid line indicates the estimated theoretical Jacobi radius ($R_J \approx 0\overset{\circ}{.}35$). The fact that $R_{90}$ is smaller than the estimated $R_J$, together with the compact half-number radius ($R_{50} = 0.^{\circ}077$), supports the interpretation that the observed population primarily samples the compact inner region of the present-day cluster.

## 4. Results

### 4.1. Lagrangian Radii and the Compactness of the Present-day Cluster

To quantitatively and robustly defend our hypothesis that the observed population constitutes a highly concentrated, dynamically overevolved core, we calculated the Lagrangian radii of our high-probability Gaia DR3 members. Based on the angular distances from the cluster center, we determined the radii enclosing 10%, 50%, and 90% of the observed members. Based on the angular distances from the cluster center, we determined the radii enclosing 10%, 50%, and 90% of the observed members (see Figure 3).

Our analysis yields $R_{10} = 0\overset{\circ}{.}025$, a half-number radius of $R_{50} = 0\overset{\circ}{.}077$, and $R_{90} = 0\overset{\circ}{.}229$. Assuming a nominal distance of ~4.0 kpc for NGC 6791 (e.g., K. Brogaard et al. 2011), these angular sizes translate to physical radii of approximately 1.7, 5.4, and 16.0 pc, respectively.

Crucially, our $R_{90}$ value (16.0 pc) remains well within the theoretical Jacobi tidal radius ($R_J$) of the cluster, which is estimated to be $\gtrsim$23 pc (E. Dalessandro et al. 2015). More importantly, the extreme compactness of $R_{50}$ (~5.4 pc) supports the view that our sample predominantly probes the compact inner regions of the present-day cluster.

In such a dense, stripped core environment, the local two-body relaxation time is drastically shorter than the cluster's extreme age of ~8 Gyr. In such a compact environment, the local two-body relaxation time is expected to be shorter than the cluster age. The inner regions are therefore likely to be dynamically well relaxed. This naturally explains the observed spatial uniformity ($p > 0.4$) between single and binary stars. This provides a plausible explanation for the observed spatial uniformity between single-star and binary/multiple candidates: long-term encounters may have mixed the surviving populations within the presently observed field, reducing any residual mass-segregation signature.

### 4.2. Confirmation of Multiplicity Dominance

Our Bayesian results indicate that rapid rotation is unlikely to be the dominant driver of the observed MS broadening. High-resolution spectroscopy has shown NGC 6791 to be chemically homogeneous ($\sigma_{[Fe/H]} < 0.02$ dex, S. Villanova et al. 2018), ruling out metallicity spread as an alternative explanation. Together with the modest measured impact of differential reddening and the spectroscopic evidence for chemical homogeneity, these results suggest that unresolved companions are likely a major contributor to the residual MS width.

### 4.3. Assessment of Candidate Unresolved Higher-order Multiples

The identification of unresolved triple systems, characterized by a magnitude excess of $\Delta G > 0.75$ mag above the single-star isochrone, requires careful differentiation from potential contamination such as chance alignments or poor astrometric solutions. We assessed the plausibility of the triple-candidate interpretation using two independent diagnostics.

First, we analyzed the Renormalized Unit Weight Error (RUWE). For our entire member sample, the median RUWE is 1.008, indicating an exceptionally clean astrometric catalog. However, the triple candidates show a distinct distribution;

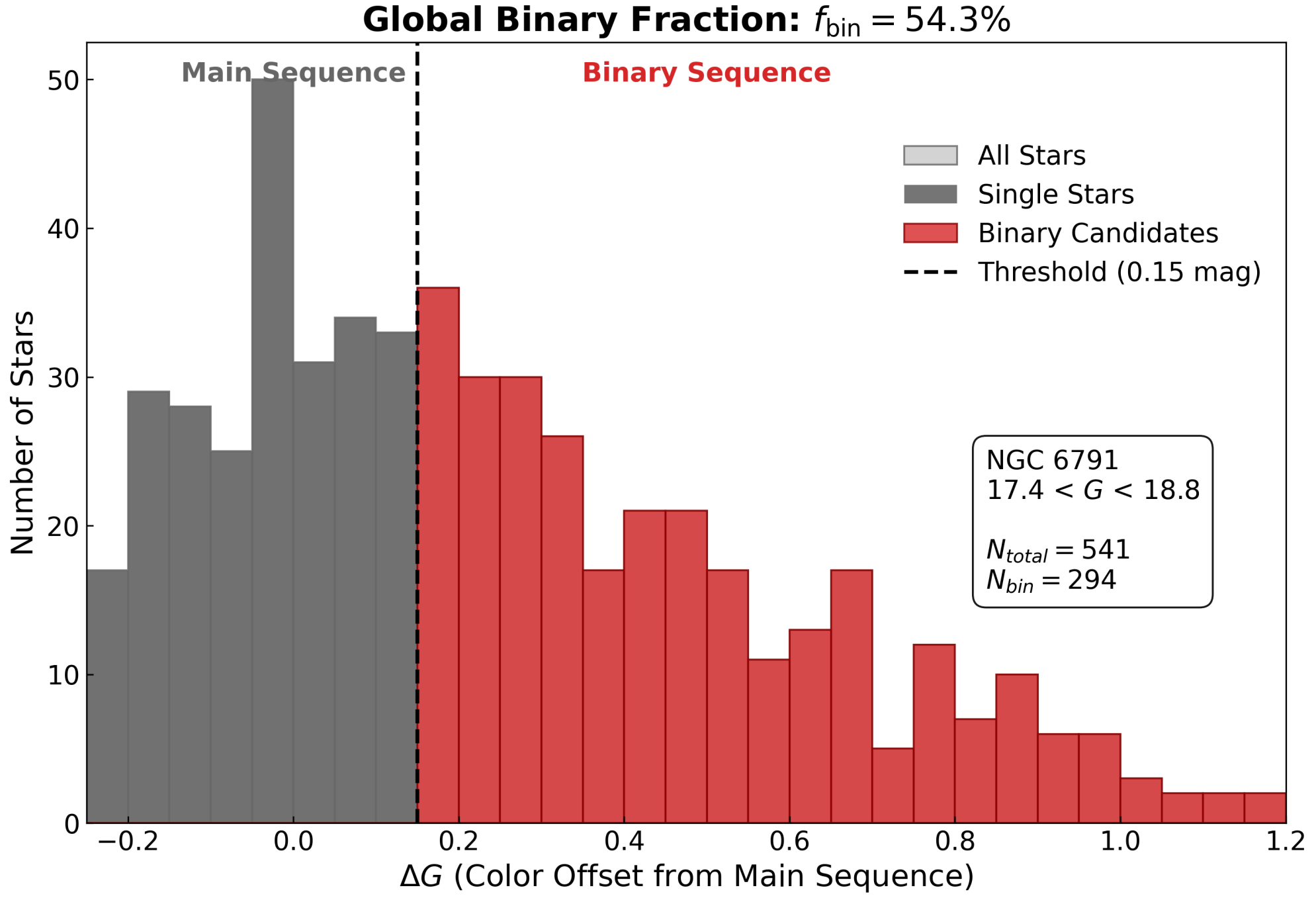

**Figure 4.** Distribution of vertical photometric offsets ($\Delta G$). The vertical dashed line at $\Delta G = 0.15$ separates singles from binaries/multiples. Note the nonuniform distribution with a significant excess of high-$q$ binaries ($0.4 < \Delta G < 0.75$) and the extended tail of triples ($\Delta G > 0.75$).

while only 1.13% of the general cluster population exceeds the standard threshold of RUWE > 1.2, a significant fraction of our identified triple-system candidates exhibit elevated RUWE values compared to the primary MS stars. Such elevations are commonly associated with non-single-star behavior, where the orbital motion of a gravitationally bound multiple system induces a "wobble" in the photocenter that cannot be adequately modeled by a standard five-parameter astrometric solution.

Second, we quantified the probability of chance alignments (photometric blends) based on the local stellar surface density. Within the core of NGC 6791, where the stellar density reaches $\rho \approx 98{,}673$ stars deg$^{-2}$, and assuming a conservative Gaia angular resolution limit of $\delta\theta \approx 0\overset{''}{.}1$, the probability of a random overlap is calculated as

$$\mathcal{P}_{\rm chance} = \pi\rho(\delta\theta)^2 \approx 2.39 \times 10^{-4}. \quad (4)$$

This expected contamination rate (0.02%) is nearly 2 orders of magnitude lower than our detected triple fraction of ~1.5%. This large discrepancy strongly suggests that random superpositions cannot account for the observed super-luminous sources, supporting their interpretation as likely hierarchical multiples. The presence of such candidates in the compact cluster core is consistent with the idea that higher-order multiples can survive and possibly be shaped by long-term few-body interactions in dynamically evolved environments.

### 4.4. An Unusually High Binary Fraction

Counting stars in the defined regions, we derive a "high-$q$" binary fraction ($q \gtrsim 0.5$) of

$$f_{\rm bin} = \frac{N_{\rm bin} + N_{\rm triple}}{N_{\rm tot}} = 54.3\% \pm 2.8\%, \quad (5)$$

where the uncertainty is estimated from bootstrap resampling. This value is higher than the typical ~20%–35% photometric binary fractions reported for many OCs and for comparable high-$q$ selections in field populations, although direct comparisons depend on the adopted mass-ratio threshold, radial coverage, and membership selection. It is consistent with a relative depletion of single stars and soft binaries in a dynamically evolved cluster. We caution that the sharp boundary at $\Delta G = 0.15$ mag is necessarily an operational photometric definition. Given the residual color scatter of the MS, some objects near this boundary may be scattered between the single-star and binary-candidate regions. The quoted fraction should therefore be interpreted as the observed high-$q$ photometric companion fraction for our adopted CMD selection rather than as a fully deconvolved intrinsic binary fraction.

### 4.5. The Hard-binary Excess

The distribution of $\Delta G$ (Figure 4) is a key diagnostic of dynamical evolution. If the intrinsic mass-ratio distribution were approximately flat, the resulting $\Delta G$ distribution, after accounting for the $q \rightarrow \Delta G$ mapping, would not be expected to show a strong concentration toward the high-$q$ end. Instead, we observe the following:

1. A dearth of systems at low offsets (excluding the single-star peak), suggesting the disruption of soft, intermediate-$q$ binaries.
2. A prominent excess in the range $0.4 < \Delta G < 0.75$, corresponding to $q \gtrsim 0.7$. This indicates a preferential retention of "hard" binaries, as predicted by Heggie's law.

A Kolmogorov–Smirnov test comparing the observed $\Delta G$ distribution (for $0.15 < \Delta G < 0.75$) to a distribution expected from a flat $q$ distribution yields a $p$-value of 0.02, confirming that the excess is statistically significant.

### 4.6. Evidence for Unresolved Triples

A striking feature of Figure 4 is the population with $\Delta G > 0.75$ mag, extending up to $\sim$1.2 mag. These stars are too bright to be simple binary systems constructed from the MS. Given the advanced age of the cluster ($\sim$8 Gyr), contamination from pre-MS stars is not expected. Furthermore, the expected contribution from optical superpositions is negligible (see Section 3.4). Thus, hierarchical multiplicity provides the most plausible explanation. Residual reddening and random photometric scatter are unlikely to account for this high-offset population. In the adopted magnitude range, the median $G$-band uncertainty is only $\sigma_G = 0.0010$ mag, and the median color uncertainty is $\sigma_{\mathrm{BP-RP}} = 0.0140$ mag. Although the residual color scatter may affect the classification of stars close to the $\Delta G = 0.15$ mag boundary, it is far too small to explain offsets exceeding the equal-mass binary limit by several tenths of a magnitude. We identify them as likely hierarchical triple systems (e.g., an inner binary plus a tertiary companion). A clearly nonnegligible population (with negligible chance-alignment contamination) suggests a dense dynamical environment where three-body interactions are frequent. The fraction of such candidates is $\sim$1.5% of the total sample, broadly consistent with the occurrence rate of hierarchical triples among solar-type field stars (A. Tokovinin 2014; M. Moe & R. Di Stefano 2017; S. S. R. Offner et al. 2023; C. Shariat et al. 2025), indicating that such systems can survive within the cluster environment.

### 4.7. Spatial Homogeneity

To test whether the binary population is more centrally concentrated than the single-star population, we compared their cumulative radial distributions within the MS analysis sample ($17.4 < G < 18.8$). We classified stars with $\Delta G \leqslant 0.15$ mag as single-star candidates and stars with $\Delta G > 0.15$ mag as binary candidates, yielding 247 single stars and 294 binaries.

A two-sample Kolmogorov–Smirnov test gives $D_{\mathrm{KS}} = 0.057$ and $p_{\mathrm{KS}} = 0.744$, indicating no statistically significant difference between the two projected radial distributions. To further quantify the expected level of stochastic variation under the null hypothesis of no intrinsic spatial segregation, we performed a Monte Carlo label-shuffling experiment with $10^4$ realizations, randomly reassigning the single/binary labels while preserving the observed sample sizes. The resulting empirical probability is $p_{\mathrm{MC}} = 0.712$.

As shown in Figure 5, the observed cumulative distributions closely follow each other over the full radial range, and the measured $\Delta\mathrm{CDF}(R)$ remains well within the null-model confidence envelopes. We therefore find no evidence that binaries are more centrally concentrated than single stars within the observed field. This result supports the interpretation that the surviving stellar population of NGC 6791 is dynamically well mixed, consistent with the picture that only the inner remnant of the original cluster remains bound today.

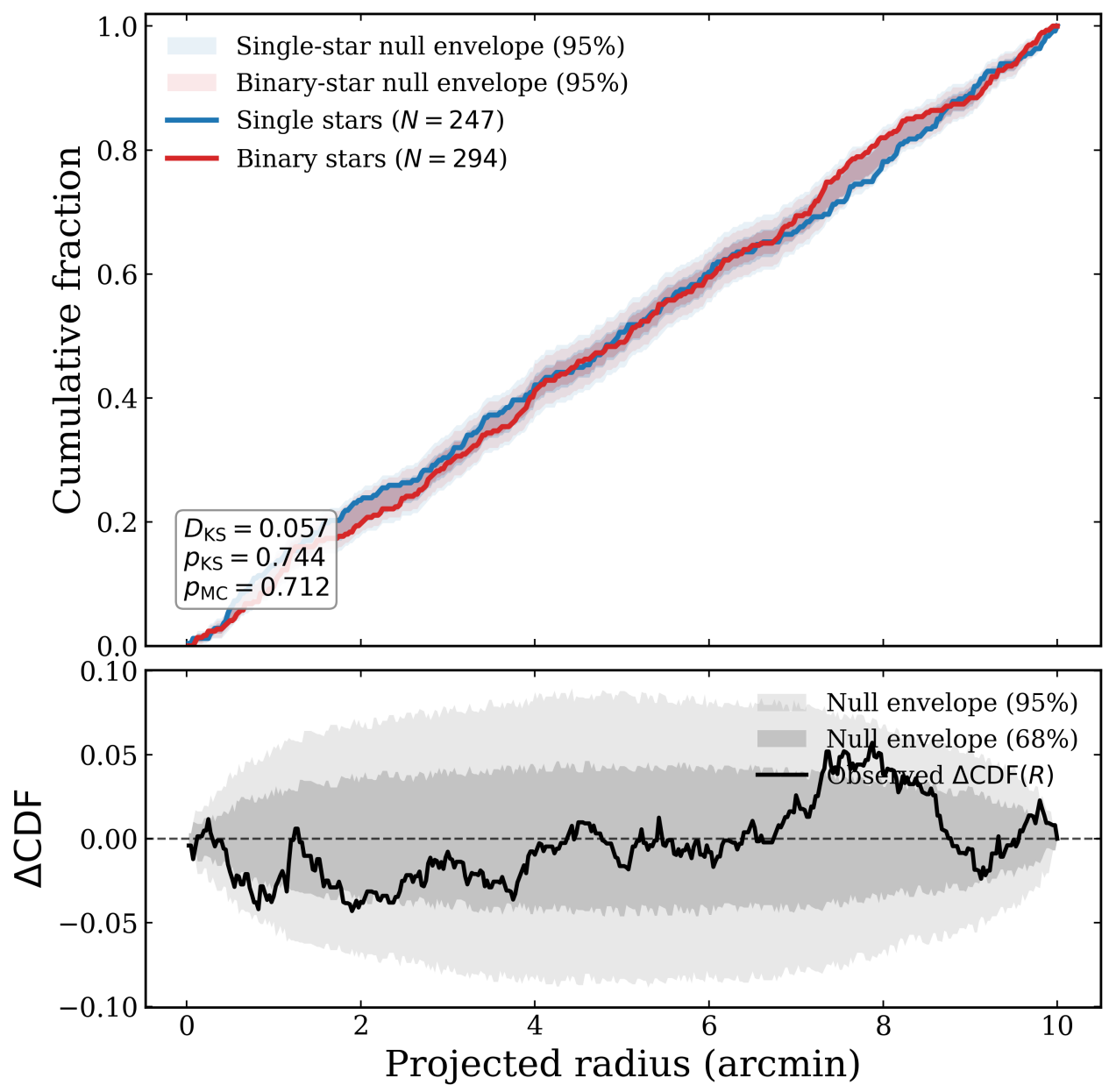

**Figure 5.** Cumulative radial distributions of single stars (blue; $N = 247$) and binary/multiple candidates (red; $N = 294$) in the MS analysis sample. Single stars are defined as objects with $\Delta G \leqslant 0.15$ mag, while binaries are defined as those with $\Delta G > 0.15$ mag. The shaded regions in the upper panel show the 95% confidence envelopes expected under the null hypothesis of no intrinsic spatial segregation, derived from $10^4$ Monte Carlo realizations in which the single/binary labels are randomly shuffled while preserving the observed sample sizes. The lower panel shows the observed difference curve, $\Delta$CDF ($R$), together with the 68% and 95% null envelopes. A two-sample Kolmogorov–Smirnov test gives $D_{\mathrm{KS}} = 0.057$ and $p_{\mathrm{KS}} = 0.744$, while the Monte Carlo test yields $p_{\mathrm{MC}} = 0.712$. The observed signal is fully consistent with the null model, indicating no statistically significant radial segregation between the single-star and binary populations within the observed field.

## 5. Discussion

The combination of a high binary fraction (54.3%), a hard-binary excess, the presence of triples, and spatial homogeneity points to a unified dynamical narrative for NGC 6791.

### 5.1. Dynamical Survivorship

The excess of high-$q$ binaries is qualitatively consistent with the preferential survival of hard systems expected from Heggie's law in a dynamically evolved dense environment. In the dense cluster core, energetic encounters disrupt "soft" (low-$q$) binaries, while "hard" binaries (high-$q$) become harder and release energy to the cluster, delaying collapse. The observed population may preferentially trace the systems that survived long-term dynamical processing: systems sufficiently tightly bound to survive long-term dynamical processing. This scenario is supported by the observed $q$ distribution, which we compared with synthetic models assuming different intrinsic $f(q)$. A model that favors high mass ratios ($q > 0.7$) reproduces the data significantly better than a flat $q$ distribution (as indicated by the Kolmogorov–Smirnov test and visual inspection of overplotted models).

### 5.2. Comparison with Binary Fractions in Other Stellar Populations

The high-$q$ companion fraction inferred for NGC 6791, $f_{\mathrm{bin}}(q \gtrsim 0.5) = 54.3\% \pm 2.8\%$, is high compared with most photometric binary fractions reported for OCs and for solar-type field stars. We note, however, that binary fractions in the literature are not always directly comparable because different studies adopt different mass-ratio limits, radial apertures, photometric filters, membership criteria, and completeness corrections. Our quoted value applies only to systems detectable above our photometric threshold, approximately $q \gtrsim 0.5$, and includes candidate higher-order multiples above the equal-mass binary sequence.

For solar-type field stars, the total multiplicity fraction is of order ∼40%–50%, but the fraction of companions above a restricted mass-ratio threshold is lower and depends on period and mass-ratio selection (D. Raghavan et al. 2010; M. Moe & R. Di Stefano 2017). In OCs, photometric binary fractions are commonly found to be ∼20%–35%, although substantial cluster-to-cluster variations exist and the binary fraction can increase toward cluster centers or in dynamically evolved systems. For example, studies of rich OCs and intermediate-age clusters have typically found lower high-$q$ binary fractions than inferred here, while some old dynamically evolved systems show enhanced central binary fractions (e.g., A. P. Milone et al. 2012; G. Cordoni et al. 2023).

Within this context, NGC 6791 lies at the high end of the observed distribution. Its large inferred high-$q$ companion fraction, together with the excess of systems near the equal-mass binary sequence, is therefore consistent with a dynamically processed population in which soft binaries and loosely bound single stars have been preferentially depleted.

### 5.3. The Stripped Core Scenario

In a relaxed cluster, equipartition of energy should cause heavy binaries to sink to the center (mass segregation). The observed spatial homogeneity is naturally accommodated if the present-day sample primarily traces the dense surviving core of the cluster. We propose that NGC 6791 has lost its outer envelope (where a larger fraction of segregated low-mass objects and single stars might once have resided) due to intense tidal stripping by the Galactic potential. What remains may be a chemically homogeneous and binary-rich inner remnant that is dynamically well mixed over the presently observed field. The lack of segregation within the observed field may reflect effective dynamical mixing, possibly aided by binary–single interactions often described as "binary burning" (E. Laplace et al. 2021). This interpretation aligns with the cluster's unusually small Jacobi radius filling factor and its high vertical orbital amplitude, which may have facilitated tidal stripping.

### 5.4. Comparison with Theoretical Models

Our findings are consistent with $N$-body simulations of old, massive clusters (e.g., J. R. Hurley et al. 2005), which predict that after several relaxation times, the surviving binary population is dominated by hard, high-$q$ systems and that the cluster core becomes well mixed. The presence of triple candidates further hints at the importance of three-body processes in shaping the present-day multiplicity.

### 5.5. The Role of Stellar Rotation and Magnetic Braking

A potential source of degeneracy in the determination of the binary fraction from the MS width is the broadening induced by stellar rotation. For young OCs, rapid rotation can shift stars redward, mimicking the photometric signature of low-mass-ratio binaries. However, NGC 6791 is one of the oldest known OCs, with an estimated age of approximately 8 Gyr (e.g., F. Grundahl et al. 2008).

According to the Skumanich-like angular momentum evolution laws (A. Skumanich 1972), solar-type stars experience significant spin-down due to magnetic braking over gigayear timescales. Modern rotational evolution models (e.g., S. P. Matt et al. 2015) predict that for stars with masses $0.85-1.1\,M_\odot$ at an age of 8 Gyr, the surface equatorial velocities should have converged to a slow-rotation regime, typically $v \sin i \lesssim 2$–$5\ \mathrm{km\,s^{-1}}$. Such low velocities result in a photometric color excess that is orders of magnitude smaller than the typical observational uncertainties of Gaia DR3 and the displacement caused by unresolved companions.

In this study, we explicitly included a rotation-broadened model in our Bayesian inference framework not as a primary explanation for the MS width, but as a rigorous null hypothesis to break the potential photometry degeneracy. Our results, which yield a Bayesian evidence factor ($\ln K$) strongly disfavoring the rotation-dominated model, are in excellent agreement with these physical expectations of magnetic braking. By quantitatively ruling out rotation, we ensure that the high binary fraction (∼54.3%) and the observed hard-binary excess are robustly attributed to the cluster's dynamical evolution rather than rotational artifacts.

### 5.6. Reconciling Dynamical Filtering with Spatial Uniformity

The observed spatial uniformity between single and binary stars ($p > 0.4$, as shown in Section 3) initially appears to contradict the hypothesis of dynamical filtering, which typically implies mass segregation. However, this lack of radial gradient can be understood within the framework of a "highly evolved stripped core."

NGC 6791 has an age of ∼8 Gyr, which is orders of magnitude longer than its typical half-mass relaxation time, $t_{\rm rh} \approx 10^8$ yr (G. Meylan 1987). In such a dynamically "overcooked" system, the cluster has likely entered a post-core-collapse phase where the inner regions are fully thermalized. The same class of long-term few-body interactions that can preferentially remove loosely bound stars and soft binaries may also help mix the surviving hard-binary population.

Furthermore, the Gaia-detected members primarily represent the dense, surviving core of a much larger progenitor. Given the strong tidal stripping experienced by NGC 6791 during its Galactic orbital evolution, the majority of the mass-segregated halo has been lost to the tidal tails. The spatial uniformity we observe is therefore not an indication of a lack of dynamical evolution, but rather a signature of a fully mixed, high-density residue where the characteristic segregation length scale now exceeds the current truncated cluster radius, $R_{\rm obs} \ll R_{\rm J}$, where $R_{\rm J}$ is the Jacobi radius.

## 6. Conclusions

Using Gaia DR3 photometry and Bayesian modeling, we have characterized the binary population of the old OC NGC 6791.

1. We find that neither metallicity dispersion nor a rotation-dominated scenario can plausibly account for the observed MS broadening; a Bayes factor of $\ln K = -3.1$ strongly favors the multiplicity model.

2. We reveal an extraordinarily high binary fraction of 54.3% ± 2.8% for systems with mass ratios $q \gtrsim 0.5$, significantly exceeding typical values for OCs and the field.
3. The mass-ratio distribution is skewed toward high-$q$ systems, with a statistically significant excess of "hard" binaries ($q \gtrsim 0.7$), accompanied by a ∼1.5% population of candidate triple systems ($\Delta G > 0.75$ mag), whose chance-alignment contamination is negligible.
4. The lack of spatial segregation is consistent with the interpretation that the present-day cluster represents a stripped inner remnant of a once more massive system, representing the dense, surviving remnant of a once more massive system that has lost its envelope through tidal stripping.

These results provide strong evidence for dynamical filtering in an old, massive cluster and demonstrate the power of precise Gaia photometry combined with Bayesian inference to uncover the imprints of long-term dynamical evolution.

## Acknowledgments

This work has used data from the European Space Agency (ESA) Gaia mission (https://www.cosmos.esa.int/gaia), processed by the Gaia Data Processing and Analysis Consortium (DPAC; https://www.cosmos.esa.int/web/gaia/dpac/consortium). We thank the developers of the `dynesty` and `jax` packages for making their tools available.

## Data and Code Availability

Python code and member star data related to this Bayesian model are available on the GitHub repository: https://github.com/chihuanbin/NGC6791. The repository includes the full nested sampling implementation, isochrone handling routines, and differential reddening correction scripts.

## Funding

This work is supported by the Scientific Research Fund of the Yunnan Provincial Department of Education (2026J0823). This work is supported by the National Natural Science Foundation of China (NSFC; 12433012, 12373097) and the Yunnan Basic Research Program (No. 202501AT070027). This work was also supported by the On-campus Construction Project of the Artificial Intelligence Laboratory, Yunnan Open University.

## Appendix A
## Differential Reddening Correction

To mitigate the broadening effects of spatially variable extinction, we implemented a correction following A. P. Milone et al. (2012). For each target star, we selected the $k = 30$ nearest neighbors in spatial coordinates ($\Delta\alpha \cos\delta$, $\Delta\delta$) that lie along the MS ridge line. The local differential reddening, $\delta E(\mathrm{BP} - \mathrm{RP})$, is calculated as

$$\delta E(\mathrm{BP} - \mathrm{RP})_i = \mathrm{median}[(G_{\mathrm{BP}} - G_{\mathrm{RP}})_{\mathrm{obs}} - (G_{\mathrm{BP}} - G_{\mathrm{RP}})_{\mathrm{RL}}]_{k=30}. \quad \text{(A1)}$$

This map was then smoothed with a Gaussian kernel of width 30″ and applied to all cluster members. Figure A1 demonstrates the effectiveness of this correction in tightening the MS.

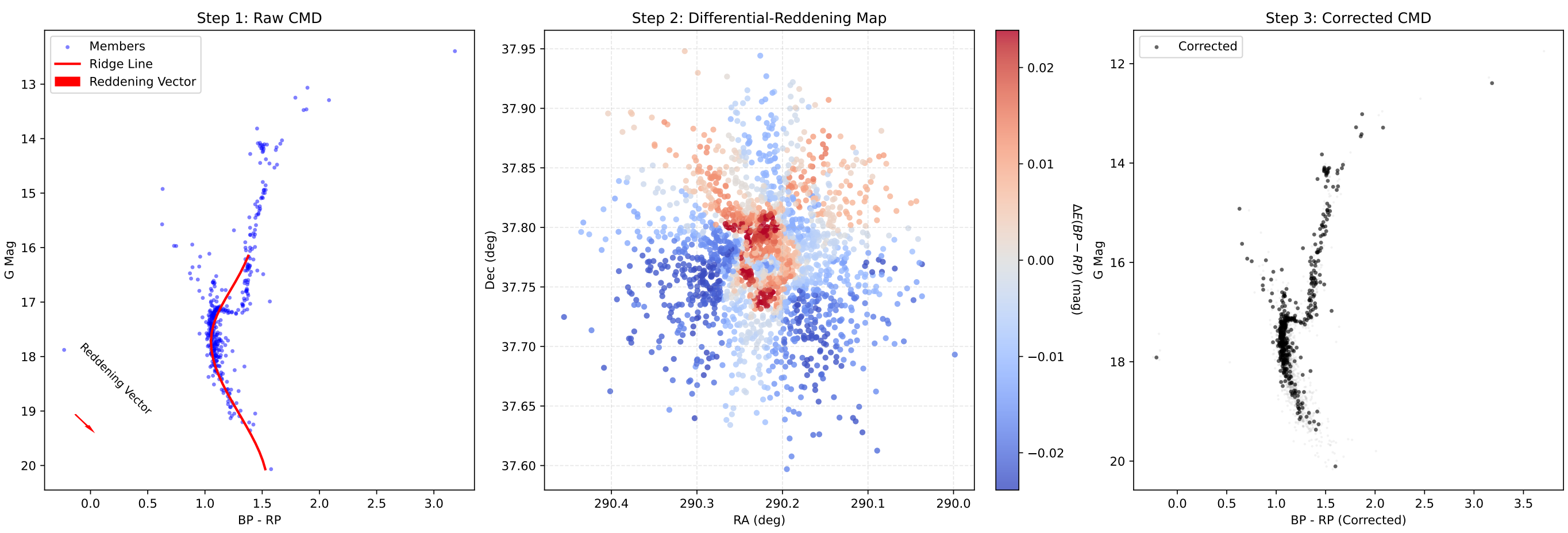

**Figure A1.** Effect of differential reddening correction. Left: raw CMD. Right: corrected CMD. The tightening of the sequence allows for precise binary discrimination.

## Appendix B
## Details of Bayesian Model Comparison

The likelihood function for a given model (rotation or no-rotation) is

$$\mathcal{L}(\boldsymbol{\theta}) = \prod_{i=1}^{N} \left[ \sum_j w_j \frac{1}{2\pi\sigma^2} \exp\left( -\frac{(c_i - c_j)^2 + (g_i - g_j)^2}{2\sigma^2} \right) \right], \tag{B1}$$

where the sum over $j$ runs over all isochrone points, $w_j$ is IMF weights, and $\sigma = 0.05$ mag accounts for photometric errors and intrinsic scatter. For the rotation model, the isochrone points are a mixture of slow and fast rotator tracks with fraction $f_{\text{fast}}$.

Nested sampling was performed using the `dynesty` package with the following settings: 300 live points, bounding "multi," sampling "rwalk," and stopping criterion "dlogz=0.5." The prior distributions were as described in Section 3.2. The results were robust to doubling the number of live points.

## ORCID iDs

Huanbin Chi (迟焕斌) https://orcid.org/0000-0001-7343-7332
Zhi Li (李志) https://orcid.org/0000-0003-0418-8461
Feng Wang (王锋) https://orcid.org/0000-0002-9847-7805
Xuefen Tian (田雪芬) https://orcid.org/0009-0003-3832-6962
Linfeng Chang (常林凤) https://orcid.org/0000-0002-8421-4561
Hongbo Liu (刘红波) https://orcid.org/0009-0008-6777-0997
Yiqin Liu (刘艺琴) https://orcid.org/0009-0009-5768-7775